% 5 points
\font\fiverm=cmr5
\font\fivebf=cmbx5
\font\fivei=cmmi5

\font\fivesy=cmsy5
% 6 points

% 7 points
\font\sevenrm=cmr7
\font\sevenbf=cmbx7
\font\seveni=cmmi7

\font\sevensy=cmsy7
% 8 points

% 9 points
\font\ninerm=cmr9
\font\ninebf=cmbx9

\font\nineit=cmmi9
\font\ninei=cmmi9 
\font\ninesy=cmsy9  
\font\nineex=cmex10
% 10 points
\font\tenrm=cmr10
\font\tenbf=cmbx10
\font\tensl=cmsl10
\font\tenit=cmmi10
\font\teni=cmmi10 
 
\font\tensy=cmsy10
\font\tenex=cmex10
% 12 points
\font\twelverm=cmr12
\font\twelvebf=cmbx12
\font\twelvesl=cmsl12
\font\twelveit=cmmi12
\font\twelvei=cmmi12
\font\twelvesy=cmsy10 scaled\magstep1

% 14 points

%
% Font family 8pt
%
 %
% Font family 9pt
%
 \def\ninepoint{%
   \normalbaselineskip=11pt
   \def\rm{\fam0\ninerm}%
   \def\it{\fam0\nineit}%
   \def\bf{\fam\bffam\ninebf}%
   \def\bi{\fam\bffam\ninebf}%
   \def\rmit{\fam0\ninerm\def\it{\fam0\nineit}}%
   \def\bfit{\fam\bffam\ninebf\def\it{\bi}}%
   \def\bsl{\fam\bffam\ninebsl}
   \textfont0=\ninerm\scriptfont0=\sevenrm\scriptscriptfont0=\fiverm
   \textfont1=\ninei\scriptfont1=\seveni\scriptscriptfont1=\fivei
   \textfont2=\ninesy\scriptfont2=\sevensy\scriptscriptfont2=\fivesy
   \textfont3=\nineex \scriptfont3=\tenex \scriptscriptfont3=\tenex
   \textfont\bffam=\ninebf\scriptfont\bffam=\sevenbf\scriptscriptfont\bffam=
     \fivebf
   \normalbaselines\rm}%
% Font family 10pt
%
 \def\tenpoint{%
   \normalbaselineskip=12pt
   \def\rm{\fam0\tenrm}%
   \def\it{\fam0\tenit}%
   \def\bf{\fam\bffam\tenbf}%
   \def\bi{\fam\bffam\tenbf}%
   \def\rmit{\fam0\tenrm\def\it{\fam0\tenit}}%
   \def\bfit{\fam\bffam\tenbf\def\it{\bi}}%
   \def\bsl{\fam\bffam\tenbsl}
   \textfont0=\tenrm\scriptfont0=\sevenrm\scriptscriptfont0=\fiverm
   \textfont1=\teni\scriptfont1=\seveni\scriptscriptfont1=\fivei
   \textfont2=\tensy\scriptfont2=\sevensy\scriptscriptfont2=\fivesy
   \textfont3=\tenex \scriptfont3=\tenex \scriptscriptfont3=\tenex
   %\textfont\bffam=\tenbf\scriptfont\bffam=\sevenbf\scriptscriptfont\bffam=
     \fivebf
   \textfont\bffam=\tenib\scriptfont\bffam=\sevenib\scriptscriptfont\bffam=
     \fiveib
   \normalbaselines\rm}%
% Font family 12pt
%

%
% Font famile 14pt

%  Define a whole menagerie of pseudo-12pt fonts
     
\font\twelverm=cmr10 scaled 1200    \font\twelvei=cmmi10 scaled 1200
\font\twelvesy=cmsy10 scaled 1200   \font\twelveex=cmex10 scaled 1200
\font\twelvebf=cmbx10 scaled 1200   \font\twelvesl=cmsl10 scaled 1200
\font\twelvett=cmtt10 scaled 1200   \font\twelveit=cmti10 scaled 1200
     
\skewchar\twelvei='177   \skewchar\twelvesy='60
     
%  Define \...point macros to change fonts and spacings consistently
     
\def\twelvepoint{\normalbaselineskip=12.4pt
  \abovedisplayskip 12.4pt plus 3pt minus 9pt
  \belowdisplayskip 12.4pt plus 3pt minus 9pt
  \abovedisplayshortskip 0pt plus 3pt
  \belowdisplayshortskip 7.2pt plus 3pt minus 4pt
  \smallskipamount=3.6pt plus1.2pt minus1.2pt
  \medskipamount=7.2pt plus2.4pt minus2.4pt
  \bigskipamount=14.4pt plus4.8pt minus4.8pt
  \def\rm{\fam0\twelverm}          \def\it{\fam\itfam\twelveit}%
  \def\sl{\fam\slfam\twelvesl}     \def\bf{\fam\bffam\twelvebf}%
  \def\mit{\fam 1}                 \def\cal{\fam 2}%
  \def\tt{\twelvett}
  \textfont0=\twelverm   \scriptfont0=\tenrm   \scriptscriptfont0=\sevenrm
  \textfont1=\twelvei    \scriptfont1=\teni    \scriptscriptfont1=\seveni
  \textfont2=\twelvesy   \scriptfont2=\tensy   \scriptscriptfont2=\sevensy
  \textfont3=\twelveex   \scriptfont3=\twelveex  \scriptscriptfont3=\twelveex
  \textfont\itfam=\twelveit
  \textfont\slfam=\twelvesl
  \textfont\bffam=\twelvebf \scriptfont\bffam=\tenbf
  \scriptscriptfont\bffam=\sevenbf
  \normalbaselines\rm}
     
%       tenpoint
     
\def\tenpoint{\normalbaselineskip=12pt
  \abovedisplayskip 12pt plus 3pt minus 9pt
  \belowdisplayskip 12pt plus 3pt minus 9pt
  \abovedisplayshortskip 0pt plus 3pt
  \belowdisplayshortskip 7pt plus 3pt minus 4pt
  \smallskipamount=3pt plus1pt minus1pt
  \medskipamount=6pt plus2pt minus2pt
  \bigskipamount=12pt plus4pt minus4pt
  \def\rm{\fam0\tenrm}          \def\it{\fam\itfam\tenit}%
  \def\sl{\fam\slfam\tensl}     \def\bf{\fam\bffam\tenbf}%
  \def\smc{\tensmc}             \def\mit{\fam 1}%
  \def\cal{\fam 2}%
  \textfont0=\tenrm   \scriptfont0=\sevenrm   \scriptscriptfont0=\fiverm
  \textfont1=\teni    \scriptfont1=\seveni    \scriptscriptfont1=\fivei
  \textfont2=\tensy   \scriptfont2=\sevensy   \scriptscriptfont2=\fivesy
  \textfont3=\tenex   \scriptfont3=\tenex     \scriptscriptfont3=\tenex
  \textfont\itfam=\tenit
  \textfont\slfam=\tensl
  \textfont\bffam=\tenbf \scriptfont\bffam=\sevenbf
  \scriptscriptfont\bffam=\fivebf
  \normalbaselines\rm}
     
%%
%%      Various internal macros
%%

{\obeylines\gdef\
{}}
\def\singlespace{\baselineskip=\normalbaselineskip}

\def\doublespace{\baselineskip=\normalbaselineskip \multiply\baselineskip by 2}

\newcount\firstpageno
\firstpageno=2
\footline={\ifnum\pageno<\firstpageno{\hfil}\else{\hfil\twelverm\folio\hfil}\fi}
\let\rawfootnote=\footnote              % We must set the footnote style
\def\footnote#1#2{{\rm\singlespace\parindent=0pt\rawfootnote{#1}{#2}}}

%%
%%      Page layout, margins, font and spacing (feel free to change)
%%
     
\hsize=6.5truein
\hoffset=0truein
\vsize=8.9truein
\voffset=0truein
\parskip=\medskipamount
\twelvepoint            % selects twelvepoint fonts (cf. \tenpoint)
\doublespace            % selects double spacing for main part of paper (cf.
                        %       \singlespace, \oneandahalfspace)
\overfullrule=0pt       % delete the nasty little black boxes for overfull box
     
%%
%%      The user definitions for major parts of a paper (feel free to change)
%%
     
\def\preprintno#1{
 \rightline{\rm #1}}    % Preprint number at upper right of title page
     
%%
%%      Various little user definitions
%%
     
\def\ref#1{Ref. #1}                     %       for inline references
                     %       ditto

          % For citation of equation numbers
        %       ditto
                     %       ditto
                     %       ditto
                   %       ditto
                   %       ditto
\def\frac#1#2{{\textstyle{#1 \over #2}}}

\def\sla{\raise.15ex\hbox{$/$}\kern-.57em}
\def\leaderfill{\leaders\hbox to 1em{\hss.\hss}\hfill}
\def\twiddle{\lower.9ex\rlap{$\kern-.1em\scriptstyle\sim$}}
\def\bigtwiddle{\lower1.ex\rlap{$\sim$}}
\def\gtwid{\mathrel{\raise.3ex\hbox{$>$\kern-.75em\lower1ex\hbox{$\sim$}}}}
\def\ltwid{\mathrel{\raise.3ex\hbox{$<$\kern-.75em\lower1ex\hbox{$\sim$}}}}
\def\square{\kern1pt\vbox{\hrule height 1.2pt\hbox{\vrule width 1.2pt\hskip 3pt
   \vbox{\vskip 6pt}\hskip 3pt\vrule width 0.6pt}\hrule height 0.6pt}\kern1pt}

\def\Fint{\rlap{$\Biggl\rfloor$}\Biggl\lceil}

\def\m@th{\mathsurround=0pt }
\def\leftrightarrowfill{$\m@th \mathord\leftarrow \mkern-6mu
 \cleaders\hbox{$\mkern-2mu \mathord- \mkern-2mu$}\hfill
 \mkern-6mu \mathord\rightarrow$}
\def\overleftrightarrow#1{\vbox{\ialign{##\crcr
     \leftrightarrowfill\crcr\noalign{\kern-1pt\nointerlineskip}
     $\hfil\displaystyle{#1}\hfil$\crcr}}}
\input psfig.sty
\singlespace
\preprintno{hep-ph/9710466}
\preprintno{CRETE-97-11}
\preprintno{UFIFT-HEP-97-1}
\vskip 2cm
\centerline{\bf MATTER CONTRIBUTIONS TO THE EXPANSION RATE OF THE UNIVERSE}
\vskip 2cm
\centerline{\bf N. C. Tsamis$^{*}$}
\vskip .5cm
\centerline{\it Department of Physics, University of Crete}
\centerline{\it Heraklion, Crete 71003, GREECE}
\vskip 1cm
\centerline{and}
\vskip 1cm
\centerline{\bf R. P. Woodard$^{\dagger}$}
\vskip .5cm
\centerline{\it Department of Physics, University of Florida}
\centerline{\it Gainesville, FL 32611, USA}
\vskip 2cm
\centerline{ABSTRACT}
\itemitem{}{\tenpoint We consider the effect of various particles 
on the cosmic expansion rate relative to that of the graviton. 
Effectively massless fermions, gauge bosons and conformally coupled 
scalars make only minuscule contributions due to local conformal 
invariance. Minimally coupled scalars can give much stronger
contributions, but they are still sub-dominant to those of gravitons
on account of global conformal invariance. Unless effectively massless 
scalar particles with very particular couplings exist, the leading 
effect on the expansion rate is furnished solely by the graviton. 
An upper bound on the mass of such scalar particles is obtained.}
\footnote{}{$^*$~~ \tenpoint e-mail: tsamis@physics.uch.gr}
\footnote{}{$^{\dagger}$~~ \tenpoint e-mail: woodard@phys.ufl.edu}

\vfill\eject
\doublespace

\centerline {\bf 1. Introduction}

In this note, we analyze the contribution of the matter sector to the 
physical rate of expansion of the universe. The pure gravitational
contribution has been considered elsewhere [1]. There, we calculated
the expectation value of the invariant element in the presence of a 
homogeneous and isotropic state which is initially free de Sitter 
vacuum:
$$\Bigl\langle 0 \Bigl\vert \; g_{\mu \nu}(t,{\vec x}) \;
dx^{\mu} dx^{\nu} \; \Bigr\vert 0 \Bigr\rangle = 
-dt^2 + {\rm a}^2(t) \; d{\vec x} \cdot d{\vec x} 
\;\; . \eqno(1)$$
We worked on the manifold $T^3 \times \Re$, using zero temperature 
quantum field theory based on the two-parameter Lagrangian:
$${\cal L}_{\rm GR} = {1 \over 16 \pi G} 
\Bigl(R - 2 \Lambda\Bigr) \; \sqrt{-g} 
\;\; . \eqno(2)$$
We inferred the physical rate of expansion from the effective Hubble 
parameter:
$$H_{\rm eff}(t) \equiv {d \over dt} \; \ln ({\rm a}) 
\;\; , \eqno(3)$$
under the assumption that the scale of inflation $M \sim 
(\Lambda / G)^{\frac14}$ is adequately below the Planck mass $M_{\rm Pl} 
\sim G^{-\frac12}$:
$$M \ltwid 10^{-3} \; M_{\rm Pl} 
\;\; . \eqno(4)$$ 
And we discovered -- at two loops -- a decrease in this rate by an 
amount which becomes non-perturbatively large at late times:
\footnote{*}{\tenpoint Throughout this note, $a(t)$ is the scale 
factor of a homogeneous and isotropic background geometry, $\kappa^2 
\equiv 16 \pi G$ is the loop counting parameter of quantum gravity 
and $3 H^2 \equiv \Lambda$ gives the bare Hubble constant.} 
$$H_{\rm eff}(t) = H \Biggl\{ 1 - \Bigl( {\kappa H \over 4 \pi} \Bigr)^4 
\; \Bigl[ \; \frac16 \; (Ht)^2 + ({\rm subdominant}) \Bigr] \;
+ \; O(\kappa^6) \Biggr\} 
\;\; , \eqno(5)$$
This may mean that the actual cosmological constant is not unnaturally
small and that our current expansion is the result of a screening 
effect whose slow onset allows a long period of inflation without the
need for new particles or severe fine tuning.

Result (5) was obtained by taking the onset of inflation to be at $t=0$
and by using perturbation theory around the classical background:
$${\rm a}_{\rm class}(t) = \exp(Ht) 
\;\; . \eqno(6)$$
It turns out that the two-loop effect drives a crucial denominator to
zero and extinguishes inflation at a time:
$$Ht \sim \Biggl( {M_{\rm Pl} \over M} \Biggr)^{\frac83} \gtwid 10^8
\;\; . \eqno(7)$$
when all higher gravitational corrections are insignificant [2].

It is natural to wonder how the back-reaction induced by matter compares
with the graviton result (5). Since the basic mechanism for any secular
effect in vacuum must be infrared, we need consider only quanta which 
are effectively massless. This is because infrared effects influence local 
observables through the coherent superposition of distant interactions 
in the past lightcone of the observer. Massive propagators oscillate 
inside the lightcone and this leads to destructive interference. 

The phrase ``effectively massless'' means that the particle's mass
$m$ results in only small distortions of its free mode functions 
over the relaxation time (7). When $m/H \ll 1$ it turns out that 
these distortions give a multiplicative factor of $a(t)$ raised
to the $-m^2/3H^2$ power. Requiring this multiplicative factor to 
be of order one gives the following bound:
$$m \ltwid M \; \Biggl( {M \over M_{\rm Pl}} \Biggr)^{\frac73} 
\;\; . \eqno(8)$$
A high inflation scale, say $M \sim 10^{16} \; {\rm GeV}$, means 
that effectively massless particles must be less than about $10^9 \;
{\rm GeV}$. Lower inflationary scales provide a more stringent bound. 
For instance, if $M$ descends all the way to $10^3~{\rm GeV}$, then 
$m \ltwid 10^{-34} \; {\rm GeV}$.
 
If the massless particle has locally conformally invariant 
interactions, then any local quantity -- when written in conformal 
coordinates -- is the same as in flat space. However, the global 
effect is nill because the conformal coordinate volume is only 
$H^{-4}$.
\footnote{*}{\tenpoint The relevant conformal parameter range is: 
$x^i \in [-\frac12 H^{-1} , \frac12 H^{-1}) \; , \; u \in [H^{-1} , 
0)$.} 
Physically, such a massless particle is unaware of the inflating 
spacetime. All the observed effectively massless fermions and gauge 
bosons are eliminated as dominant contributors to $H_{\rm eff}(t)$ 
because of the aforementioned argument. The same is true for 
conformally coupled scalars.  

Minimally coupled scalar particles could have a leading influence on 
$H_{\rm eff}(t)$ and it is the purpose of this note to evaluate their 
effect. The first non-trivial diagrams occur at two loops -- see, for 
instance, Fig.~1a. Although they should naively contribute as strongly 
as the graviton, explicit calculation shows that they are in fact 
subleading (Section 2). Section 3 explains this as a consequence of 
{\it global} conformal invariance. In addition to its phenomenological 
significance, this result provides a severe test on the master 
integration programs used in the graviton and scalar cases. 

\vskip 1.3cm

\centerline{\psfig{figure=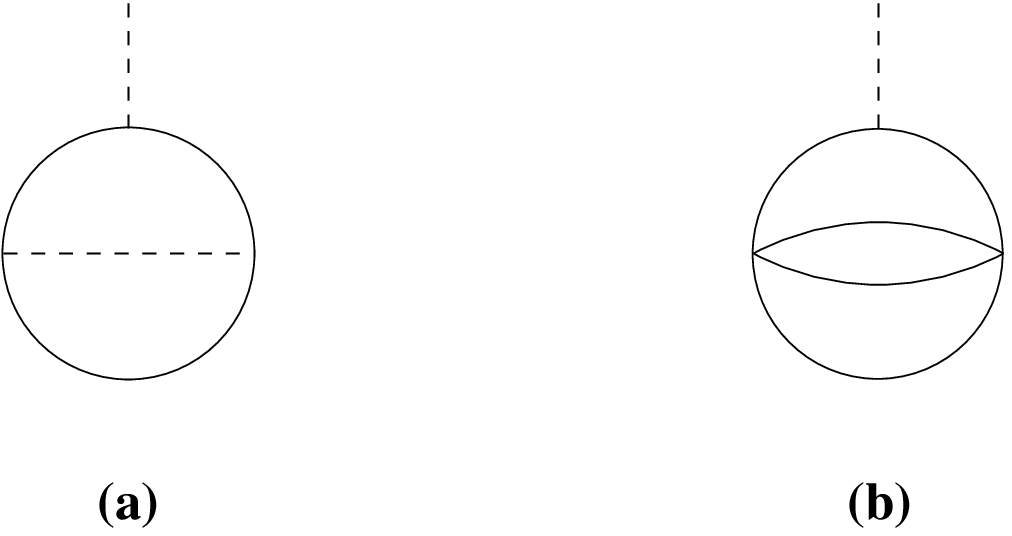,width=7truecm,angle=-90}}

\vskip 0.2cm

{\bf Fig.~1:} {\ninepoint Various contributions to 
$H_{\rm eff}(t)$. Gravitons reside on segmented lines, scalar 
fields
\vskip -18pt \noindent \hglue 2.40truecm
on solid lines.}

\vskip 0.5cm

The only exception is effectively massless scalars with non-derivative
self-interactions. (See, for example, Fig.~1b.) These are considered
in Section 4. Although they can contribute more strongly than gravitons,
the effect is always to slow inflation.

\vfill\eject
\centerline {\bf 2. The Result}

The complete Lagrangian ${\cal L}$ to be studied consists of the 
gravitational part ${\cal L}_{\rm GR}$ -- given by (2) -- and the
matter part ${\cal L}_{\rm SC}$ representing a massless minimally
coupled scalar field:
$${\cal L} = {\cal L}_{\rm GR} + {\cal L}_{\rm SC}  
\;\; , \eqno(9a)$$
$${\cal L}_{\rm SC} = - \frac12 \sqrt{-g} \; g^{\mu \nu} \;
\partial_{\mu} \phi \; \partial_{\nu} \phi 
\;\; . \eqno(9b)$$ 
As in the purely gravitational case, it is most convenient to employ
the ``open-conformal'' set of coordinates:
$$-dt^2 + {\rm a}^2_{\rm class}(t) \; d{\vec x} \cdot d{\vec x} 
= \Omega^2 \Bigl(-du^2 + d{\vec x} \cdot d{\vec x}\Bigr) 
\;\; ,\eqno(10a)$$
$$\Omega \equiv (H u)^{-1} = \exp(H t) 
\;\; , \eqno(10b)$$
and to organize perturbation theory in terms of the quantum fields 
$\phi$ and $\psi_{\mu \nu}$:
$$g_{\mu \nu} \equiv \Omega^2 \; {\widetilde g}_{\mu \nu} \equiv 
\Omega^2 \; \Bigl(\eta_{\mu \nu} + \kappa \psi_{\mu \nu}\Bigr) 
\;\; . \eqno(11)$$
Note that the conformal time $u$ is inverted with respect
to the co-moving time $t$. The onset of inflation at $t = 0$ corresponds
to $u = H^{-1}$ while the infinite future at $t = +\infty$ corresponds
to $u = 0$. 

We shall compute the scalar contributions to the amputated expectation 
value of the pseudo-graviton field in the presence of free de Sitter 
vacuum. By using the manifest homogeneity and isotropy of the theory 
and the initial state, we have expressed the expectation value in 
terms of two functions $a(u)$ and $c(u)$:
\footnote{*}{\tenpoint The gauged-fixed kinetic operator 
$D_{\mu \nu}^{~~\rho \sigma}$ is most conveniently expressed in terms 
of the kinetic operator ${\rm D}_A \equiv \Omega (\partial^2 + 
\frac2{u^2}) \Omega$ for a massless, minimally coupled scalar and the 
kinetic operator ${\rm D}_B = {\rm D}_C \equiv \Omega \> \partial^2 
\Omega$ for a conformally coupled scalar: 
$$D_{\mu \nu}^{~~ \rho \sigma} \equiv \Bigl[
\frac12 {\overline \delta}_{\mu}^{~(\rho} \; 
{\overline \delta}_{\nu}^{~\sigma)} - 
\frac14 \eta_{\mu \nu} \; \eta^{\rho\sigma} - 
\frac12 \delta_{\mu}^{~0} \; \delta_{\nu}^{~0} \; 
\delta_0^{~\rho} \; \delta_0^{~\sigma} \Bigr] {\rm D}_A + 
\delta_{(\mu}^{~~0} \; {\overline \delta}_ {\nu)}^{~~(\rho} \; 
\delta_0^{~\sigma)} \; {\rm D}_B + 
\delta_{\mu}^{~0} \; \delta_{\nu}^{~0} \; 
\delta_0^{~\rho} \; \delta_0^{~\sigma} \; {\rm D}_C 
\;\; , $$
where the bar above a symbol means that its zero component is
projected out.}
$$D_{\mu \nu}^{~~\rho \sigma} \; \Bigl\langle 0 \Bigl\vert \;
\kappa \psi_{\rho \sigma}(x) \; \Bigr\vert 0 \Bigr\rangle =
a(u) \; {\overline \eta}_{\mu \nu} + c(u) \;
\delta^0_{~\mu} \delta^0_{~\nu} 
\;\; . \eqno(12)$$
Although (12) is not gauge invariant, it is quite simple to 
obtain from it the effective Hubble constant $H_{\rm eff}(t)$ 
which is a genuine observable and the quantity of physical 
interest [1,3]. 

The complete set of Feynman rules for the purely gravitational 
part can be found elsewhere [1] and will not be presented here. 
The scalar propagator consists of a ``normal'' and a ``logarithmic''
part:
$$\eqalignno{ 
i \Delta (x;x') \approx 
\; {H^2 \over 8 {\pi}^2} \; \Biggl\{ \; 
&{2u'u \over {\Delta x}^2 - {\Delta u}^2 + 
2 i \epsilon \vert {\Delta u} \vert + \epsilon^2} \cr 
&- \ln\Bigl[H^2 \Bigl({\Delta x}^2 - {\Delta u}^2 + 
2 i \epsilon \vert {\Delta u} \vert + \epsilon^2\Bigr)\Bigr] 
\; \Biggr\} \;\; , &(13) \cr}$$
where ${\Delta x} \equiv \Vert {\vec x}' - {\vec x} \Vert$ and 
${\Delta u} \equiv u' - u$.
\footnote{*}{\tenpoint This form of the propagator is obtained by 
turning the mode sum on $T^3$ into an integral -- an excellent 
approximation since the propagator is only needed for small conformal 
coordinate separations [3].} 
To regulate the ultraviolet sector, the mode cutoff $\epsilon$ 
which appears in the propagators has been used throughout [1,3].
The relevant graviton-scalar interactions are:
$${\cal L}_{\rm INT}^{(1,2)} =
\kappa \Omega^2 \; \Bigl\{ \;
-\frac14 \eta^{\mu \nu} \; \psi \; \phi_{, \mu} \; \phi_{, \nu}
+\frac12 \; \psi^{\mu \nu} \; \phi_{, \mu} \; \phi_{, \nu}
\; \Bigr\} \;\; , \eqno(14a)$$
$$\eqalignno{ 
{\cal L}_{\rm INT}^{(2,2)} =
\kappa^2 \Omega^2 \; \Bigl\{ \;
&\frac14 \psi \; \psi^{\mu \nu} \; \phi_{, \mu} \; \phi_{, \nu}
-\frac12 \psi^{\mu}_{~\alpha} \; \psi^{\alpha \nu} \; 
\phi_{, \mu} \; \phi_{, \nu}
-\frac1{16} \eta^{\mu \nu} \; \psi^2 \; 
\phi_{, \mu} \; \phi_{, \nu} \cr
&+\frac18 \eta^{\mu \nu} \; \psi^{\alpha}_{~\beta} \; 
\psi^{\beta}_{~\alpha} \; \phi_{, \mu} \; \phi_{, \nu}
\; \Bigr\} \;\; . &(14b) \cr}$$
These interactions have the same generic form as those of pure gravity
-- two derivatives with a factor of $\Omega^2$. Since the scalar 
propagator also has the same form as that of the graviton, one might 
expect the back-reaction from the scalar to be of the same strength 
as that of the graviton.

If quantum corrections are to exceed the classical result for
$H_{\rm eff}(t)$ at late times, $a(u)$ or $c(u)$ must grow faster
than $u^{-4}$ as $u \rightarrow 0^+$ [3]. With these interactions
all diagrams in perturbation theory that contribute to (12) are 
subject to a maximum growth of $u^{-4}$ times powers of $\ln(H u)$ 
[3,4]. If the effect is to be interesting, such logarithmic terms 
must be present since pure $u^{-4}$ behavior simply renormalizes 
$\Lambda$. The two sources of logarithmic terms are:

\noindent
(i) The integration over the interaction vertices of the theory. This
is the classic source of an infrared effect -- access to an arbitrarily 
large invariant volume in the past lightcone of the observation. 

\noindent
(ii) The ``logarithmic'' piece of a propagator. This source is particular
to de Sitter spacetime and reflects the increasing correlation of the 
vacuum at constant invariant separation. 

\vskip 1.3cm

\centerline{\psfig{figure=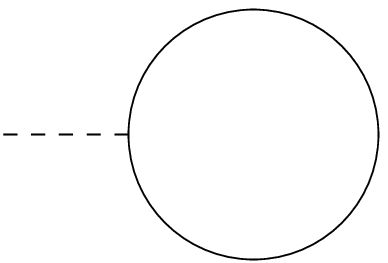,width=2cm,angle=-90}}

\vskip -0.3cm 

{\bf Fig.~2:} {\ninepoint One-loop scalar contribution to 
$H_{\rm eff}(t)$.} 

\vskip 0.5cm

Both sources are absent in the first scalar diagram that contributes 
to (12) beyond tree order (see Fig.~2). Amputation fixes the single
interaction vertex at $(u, \vec x)$ and one obtains only the coincidence
limit of derivatives of the scalar propagator at this point. In fact, 
the maximum growth we can obtain is $u^{-2}$. Therefore, we must go
to the two-loop diagrams of Fig.~3 to identify the first potentially
relevant graphs in the infrared. Starting in reverse order, diagrams 
(3d-e) are entirely canceled by the counterterms needed to renormalize 
their coincident lower loops.\footnote{*}{Not shown is the
ultra-local tadpole formed from the $\psi^3\partial\phi\partial\phi$
vertex, which must also be renormalized away.} Of the remaining graphs, 
one free interaction vertex exists in (3c) and two in (3a-b), whereas all 
three graphs can have up to one undifferentiated graviton propagator.

Our counting indicates -- in precise analogy with the pure graviton
diagrams -- that the graphs (3a-c) should contribute terms of order
$u^{-4} \ln^2(Hu)$ at late physical times. These terms would lead to 
contributions to $H_{\rm eff}(t)$ that are equal in strength to the 
pure gravitational ones (5) and, consequently, would modify the 
expansion rate of the universe.

\vskip 1.3cm

\centerline{\psfig{figure=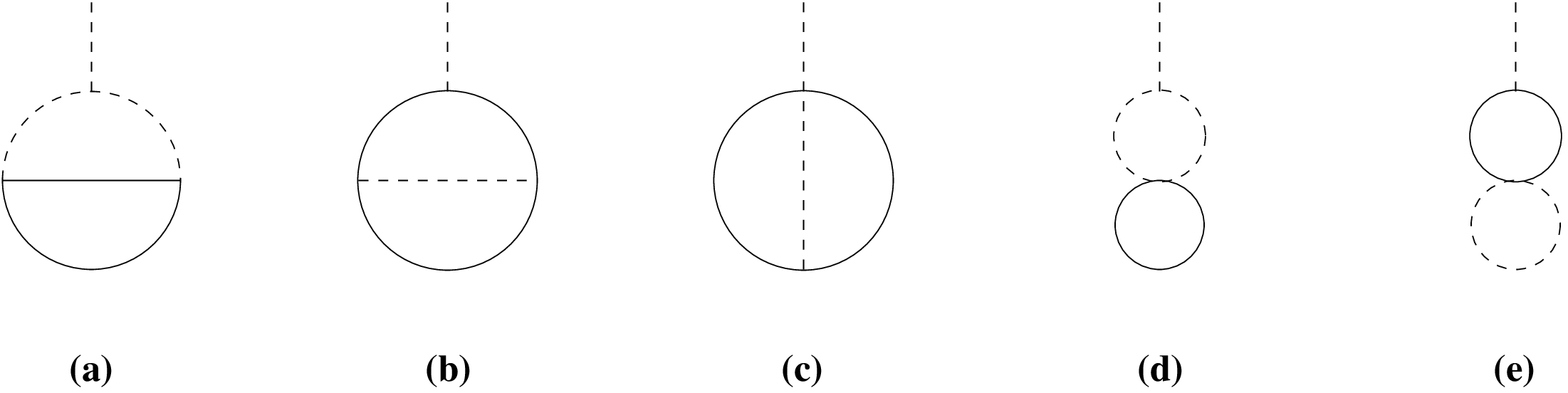,width=15cm,angle=-90}}

\vskip -0.3cm 

{\bf Fig.~3:} {\ninepoint Two-loop scalar contributions to 
$H_{\rm eff}(t)$.} 

\vskip 0.5cm

The actual calculation was performed using the same techniques as with
the more complicated pure gravitational case. The symbolic manipulation
program Mathematica [5] was used throughout.
\footnote{*}{\tenpoint Copies of the computer programs employed and the 
intermediate expressions they generated can be made available. Both the 
tensor algebra and the loop integrations used programs identical to 
those in the pure graviton case.} 
Acting the derivatives from interactions on the propagators produces 
many terms. Some of these integrate to contribute to the coefficient 
functions $a(u)$ and $c(u)$ at order $u^{-4} \ln^2(Hu)$ --- the same as 
gravitons. However, the net result is sub-dominant to that of pure 
gravitation, {\it for each of the three diagrams separately.} In fact, 
this cancellation was found to occur whenever one sums over all the pure 
graviton interactions and the various terms which come from {\it any 
combination} of the six scalar-graviton interactions given in (14a) 
and (14b).

\vskip 1cm
\centerline {\bf 3. The Justification}

The sub-dominance of diagrams involving any combination of scalar-graviton
interactions suggests that the cause is a symmetry of the scalar action 
which is not shared by the gravitational action. The natural candidate 
is global conformal invariance, whose action on the fields can be given
in terms of a spacetime constant parameter $s$:
$$g_{\mu \nu}^{'}(x) = s^2 \; g_{\mu \nu}(x) \quad , \quad
\phi^{'}(x) = s^{-1} \; \phi(x) 
\;\; . \eqno(15)$$
The scalar Lagrangian is invariant under (15) while the gravitational 
Lagrangian is not. We will see how this, combined with dimensional 
analysis and a few pieces of field theory lore suffice to explain the 
cancellation.

We seek to understand why the amputated 1-point function acquires no
terms of the form $u^{-4} \ln^2(u)$ from diagrams which involve a
scalar loop. Since $u$ is dimensionful, each factor of $\ln(u)$ must
be paired with the logarithm of some other dimensionful parameter. 
Before renormalization there are only two such parameters: the Hubble 
constant $H$ and the ultraviolet regularization parameter $\epsilon$. 
Factors of $\ln(H)$ enter from undifferentiated propagators (13), 
which must be {\it graviton} lines since all scalar fields in 
${\cal L}_{\rm SC}$ carry derivatives.
\footnote{*}{\tenpoint There can be no factors of $\ln(H)$ from the 
$H^{-1}$ limit on the conformal time integrations because power 
counting shows that the integrands converge for large conformal 
time [1].} 
Two-loop diagrams which contribute to the amputated 1-point function 
can have at most one such undifferentiated propagator [1]. Factors of 
$\ln(\epsilon)$ come from logarithmic ultraviolet divergences, of which 
there can be at most two at two loops. The two possibilities are 
therefore a double logarithmic ultraviolet divergence or else a single 
logarithmic ultraviolet divergence combined with an undifferentiated 
graviton propagator.

\vskip 1.3cm

\centerline{\psfig{figure=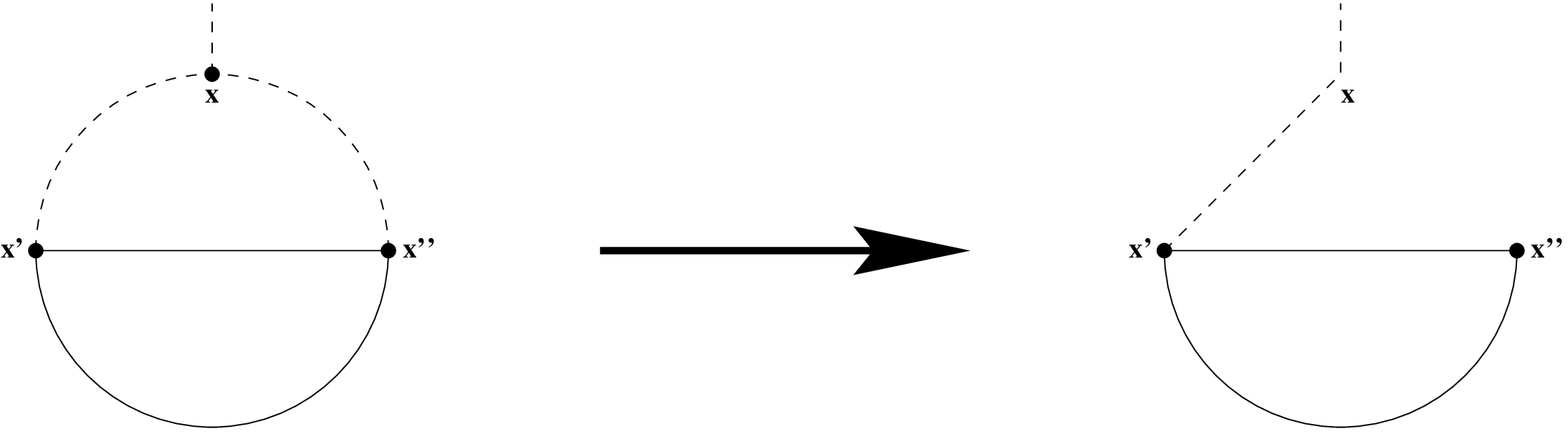,width=11truecm,angle=-90}}

\vskip 0.2cm

{\bf Fig.~4:} {\ninepoint Upon replacing the logarithmic piece of the
undifferentiated propagator -- in this case $i\Delta(x'' ; x)$ -- 
\vskip -18pt \noindent \hglue 2.40truecm
with a constant, the two-loop diagram becomes one-loop.}

\vskip 0.5cm

In either of the aforementioned possibilities, the scalar loop must
contribute a factor of $\ln(\epsilon)$. This is obvious for the double
logarithm since then each loop must be logarithmically divergent. The
statement is also true when an undifferentiated graviton propagator
contributes a factor of $\ln(H)$ because this removes any spacetime
dependence from the associated graviton line, effectively cutting it
-- see Fig.~4. But then the factor of $\ln(\epsilon)$ must come from 
the only intact loop: that of the scalar.

\vskip 1.3cm

\centerline{\psfig{figure=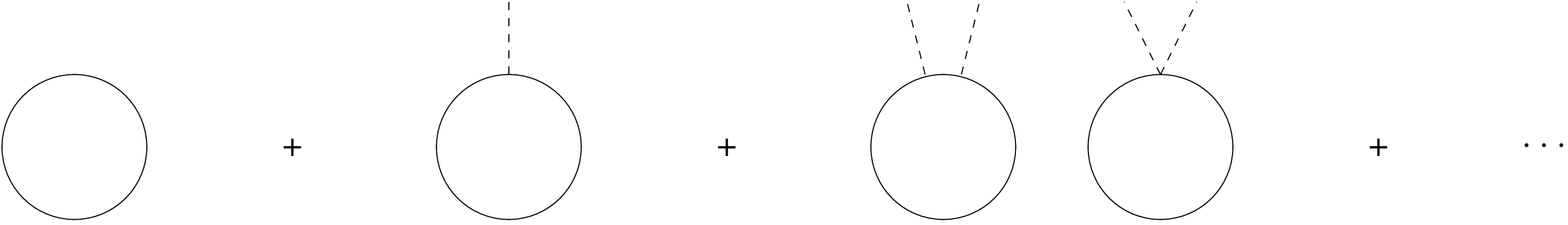,width=15cm,angle=-90}}

\vskip -0.3cm 

{\bf Fig.~5:} {\ninepoint The diagrammatic expansion of the effective 
action $\Gamma_{\rm SC}$.} 

\vskip 0.5cm

The natural vehicle for analyzing such issues is the gravitational
effective action induced by the scalar, $\Gamma_{\rm SC}[g]$ -- see 
Fig.~5. The amputated 1-point function can be expressed in terms 
of this quantity by performing the Gaussian functional integration 
over the scalar field:
\footnote{*}{\tenpoint In the interest of clarity we have suppressed 
the gauge fixing paraphernalia, as well as the forward and backwards 
evolving fields needed to give an expectation value rather than an 
``in''-``out'' amplitude.}
$$\eqalignno{ 
D_{\mu \nu}^{~~\rho \sigma} \; \Bigl\langle 0 \Bigl\vert \;
\kappa \psi_{\rho \sigma}(x) \; \Bigr\vert 0 \Bigr\rangle 
&= \Fint [d\psi] [d\phi] \;
D_{\mu \nu}^{~~\rho \sigma} \; \kappa \psi_{\rho \sigma}(x) \;
\exp \Bigl\{ i {\cal S}_{\rm GR} + i {\cal S}_{\rm SC} \Bigr\}
\;\; , &(16a) \cr
&= \Fint [d\psi] \;
D_{\mu \nu}^{~~\rho \sigma} \; \kappa \psi_{\rho \sigma}(x) \;
\exp \Bigl\{ i {\cal S}_{\rm GR} \Bigr\} \; 
\exp \Bigl\{ i \Gamma_{\rm SC} \Bigr\}
\;\; . &(16b) \cr}$$
The resulting formalism is that of a purely gravitational theory 
whose action is that of classical gravity, $S_{\rm GR}$, plus 
$\Gamma_{\rm SC}$. Two-loop contributions involving the scalar 
consist of one-loop diagrams in this purely gravitational theory, 
where a single interaction comes from $\Gamma_{\rm SC}$ and the 
rest from $S_{\rm GR}$ -- see Fig.~6.

\vskip 1.3cm

\centerline{\psfig{figure=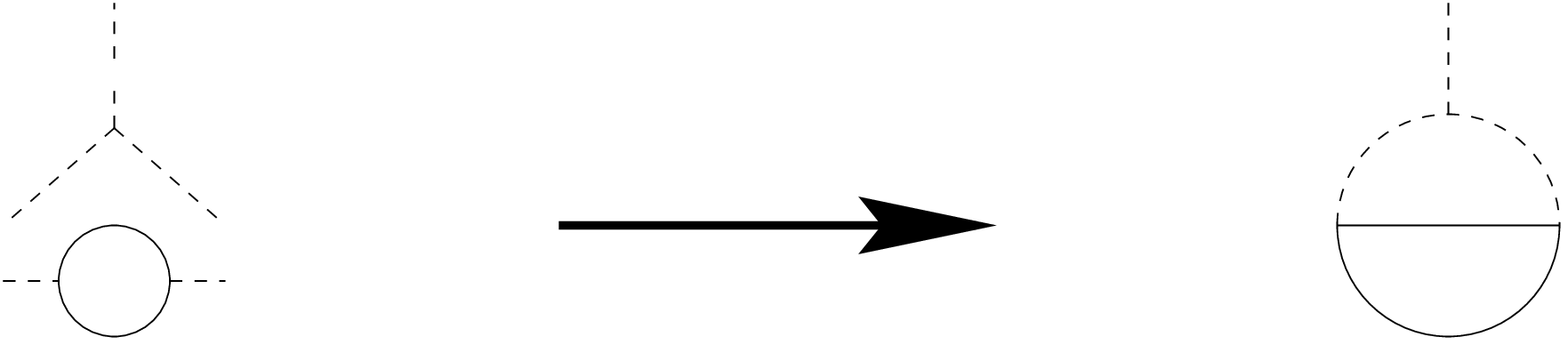,width=12cm,angle=-90}}

\vskip -0.3cm 

{\bf Fig.~6:} {\ninepoint Construction of a typical two-loop scalar 
contribution to $H_{\rm eff}(t)$.} 

\vskip 0.5cm

The ultraviolet divergences of $\Gamma_{\rm SC}$ multiply local
invariant functions of the metric, as usual. Although our mode
cut off $\epsilon$ breaks global conformal invariance, the result
of a global conformal transformation is simply to rescale $\epsilon$. 
It follows that the local invariants which multiply $\ln(\epsilon)$ 
must possess global conformal invariance. Only two such terms exist 
in four dimensions:
$$C^2 \; \sqrt{-g} \qquad , \qquad R^2 \; \sqrt{-g}
\;\; . \eqno(17)$$
The first of these is actually invariant under {\it local} conformal 
transformations, so it has no dependence upon the conformal factor
$\Omega = (Hu)^{-1}$. We also know that the expansion of its integral 
in powers of the pseudo-graviton field begins at quadratic order. The 
second term in (17) has neither of these properties: it depends upon 
$\Omega$ and the expansion of its integral begins at linear order in 
$\psi$ . One consequence of this is that $\Gamma_{\rm SC}$ cannot 
actually have a term involving $\ln(\epsilon)$ times this second term. 
If it did, the one-loop diagram of Fig.~2 would possess a logarithmic
divergence, which it does not.

Now consider one-loop contributions to the amputated 1-point function 
which come from a single insertion of an interaction from 
$C^2 \sqrt{-g}$, plus the possibility of a single interaction vertex
from ${\cal L}_{\rm GR}$ at the fixed observation point $x^{\mu}$.
These diagrams can indeed give a factor of either $\ln(\epsilon)$ or 
$\ln(H)$, {\it but without the essential multiplicative factor of} 
$u^{-4}$. The $C^2 \sqrt{-g}$ vertex is entirely independent of the
conformal time, and the vertices from ${\cal L}_{\rm GR}$ can 
contribute at most a factor of $u^{-3}$. Propagators give only {\it 
positive} powers of conformal time. We can get higher inverse powers 
from lower ones by decomposing by partial fractions terms which 
produce ultraviolet divergences:
$${1 \over u' (u' - u - i \epsilon)^2} = 
{1 \over u (u' - u - i \epsilon)^2} 
- {1 \over u^2 (u' - u - i \epsilon)} 
+ {1 \over u^2 u'} 
\;\; , \eqno(18)$$
{\it but only if the integrand already contains at least one inverse 
power of the conformal time which is being integrated.} It cannot get
{\it any} powers from the insertion, propagators give only positive
powers, and the possible vertex from ${\cal L}_{\rm GR}$ must be 
external. 

We can therefore exclude the possibility of two-loop scalar 
contributions of strength $u^{-4} \ln^2(Hu)$. In addition to
understanding why these terms can {\it not} come from free scalars
we have uncovered the reason why they {\it can} come from gravitons:
${\cal L}_{\rm GR}$ is not invariant under global conformal
transformations. We should also note that there are presummably 
non-zero scalar contributions at order $u^{-4} \ln(Hu)$ although 
we did not compute them. 

\vskip 1cm
\centerline {\bf 4. The Scalar Self-Interactions}

For completeness we consider the case of a self-interacting scalar
which somehow avoids developing an unacceptably large mass. Suppose
we add an $N$-point self-interaction:
$$- \frac1{N!} \; \lambda \; \sqrt{-g} \; \phi^N
\;\; , \eqno(19)$$
to the scalar Lagrangian (9b). There are two basic diagrams that must
be considered at the first non-trivial order, both of which involve 
the vertices:
$${\cal L}_{\rm INT}^{(0,N)} = 
- \frac1{N!} \; \lambda \Omega^4 \; \phi^N
\qquad ; \qquad
{\cal L}_{\rm INT}^{(1,N)} =
- \frac1{2N!} \; \kappa \lambda \Omega^4 \; \psi \; \phi^N
\;\; . \eqno(20)$$
The two diagrams can be seen in Fig.~7 and their respective 
contributions to the amputated expectation value (12) are:
$$\eqalignno{ 
I_{\rm (a)}^{\mu \nu} &=
\frac{i}{2(N-1)!} \; \kappa^2 \lambda^2 \Omega^2 
\int d^4x' \; {\Omega'}^4 \; \int d^4x'' \; {\Omega''}^4 \;
\Bigl( -\frac14 \eta^{\mu \nu} \; \eta^{\rho \sigma}
+\frac12 \eta^{\mu \rho} \; \eta^{\nu \sigma} \Bigr) \cr 
&\qquad\qquad\qquad\qquad\qquad
\Bigl[ \partial_{\rho} \; i\Delta(x;x') \Bigr] \;
\Bigl[ {\partial_{\sigma}} \; i\Delta(x;x'') \Bigr] \;
\Bigl[ i\Delta(x';x'') \Bigr]^{N-1} 
\;\; , \qquad &(21a) \cr
I_{\rm (b)}^{\mu \nu} &=
-\frac{i}{2N!} \; \kappa^2 \lambda^2 \Omega^4
\int d^4x' \; {\Omega'}^4 \; 
\eta^{\mu \nu} \; \Bigl[ i\Delta(x;x') \Bigr]^N 
\;\; , &(21b) \cr}$$
where $x$ is the observation event while $x'$ and $x''$ are the 
locations of the interaction vertices that must be integrated.

\vskip 1.3cm

\centerline{\psfig{figure=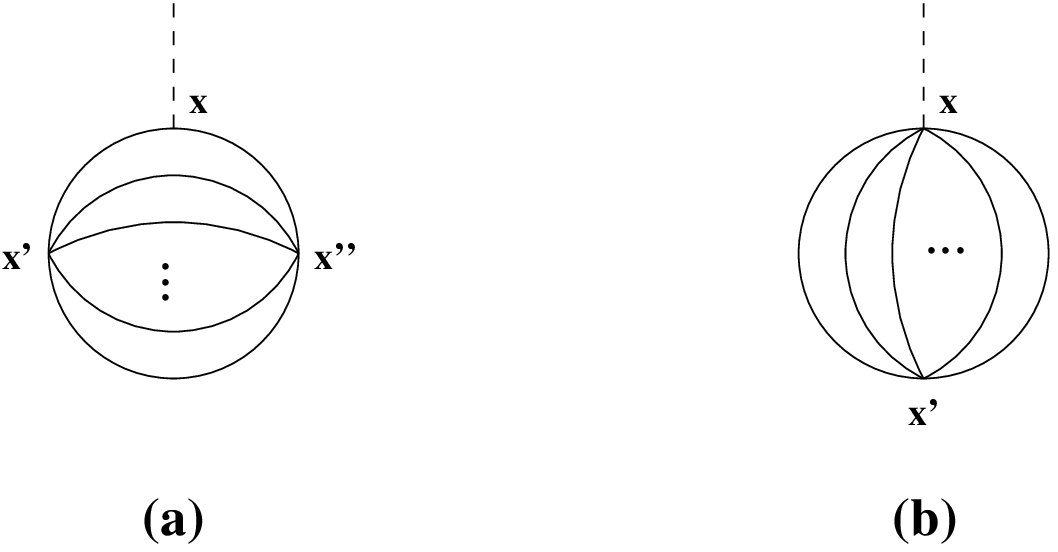,width=7truecm,angle=-90}}

\vskip -0.3cm

{\bf Fig.~7:} {\ninepoint The first basic scalar contributions 
to $H_{\rm eff}(t)$ due to an $N$-point scalar self-interaction.} 

\vskip 0.5cm

The leading contribution for diagram 7b comes entirely from the
logarithmic part of the propagator. It is easily computed to be:
$$I_{\rm (b)}^{\mu \nu} \; = \; 
-\frac16 \; {\kappa^2 \lambda^2 H^{2N-8} \over (2\pi)^{2N-2} N!} \;
{\eta_{\mu \nu} \over u^4} \; [-\ln(Hu)]^N \;
+ \; {\rm (subdominant)} 
\;\; . \eqno(22)$$
Diagram 7a is sub-dominant; it can give at most $(N-1)$ logarithms.
The effect on the expansion rate is:
$$H_{\rm eff}^{\rm SC}(t)= H \Biggl\{
1 - {\kappa^2 \lambda^2 H^{2N-6} \over (2 \pi)^{2N-2} N!} 
\Bigl[ \; \frac1{36} \; (Ht)^N 
+ \; {\rm (subdominant)} \; \Bigr] \; + \; ... \Biggr\} 
\;\; , \eqno(23)$$
Note that the effect is to slow inflation for all values of $N$.
However unlikely it is to have light, self-interacting scalars, we
need not worry that they prevent screening. They can only add to
the effect already provided by gravitons.

Note also that we get a non-zero effect even for $N=4$, when the 
scalar action has global conformal invariance. The argument of the 
previous section does not apply because non-derivative interactions 
allow the survival of many more factors of $\ln(H)$ from 
undifferentiated propagators. The integration over $u'$ also diverges 
for large conformal times, which gives another factor of $\ln(H)$.

\vfill\eject
\centerline {\bf 5. Epilogue}

The special thing about gravitons is their combination of masslessness
with an intrinsic scale which breaks conformal invariance. This feature 
is what allows them to screen the cosmological constant. We have shown 
that even global conformal invariance results in the absence of leading 
order contributions from minimally coupled scalars which lack 
self-interactions. The addition of non-derivative self-interactions 
allows minimally coupled scalars to slow inflation as much or more than
gravitons, but only if they can somehow satisfy our bound (8) for 
remaining effectively massless.

\vskip 1cm
\centerline{ACKNOWLEDGEMENTS}

We should like to thank the University of Crete and the Institute for
Fundamental Theory at the University of Florida for their hospitality 
during the execution of this project. This work was partially supported 
by DOE contract DE-FG02-97ER41029, by NSF grant 94092715, by EU grants 
940621 and 961206, by NATO grant 971166, and by the Institute for 
Fundamental Theory.

\vskip 1cm
\centerline{REFERENCES}

\item{[1]} N. C. Tsamis and R. P. Woodard, {\sl Nucl. Phys.} {\bf B474} 
(1996) 235.
\hfill\break
N. C. Tsamis and R. P. Woodard, {\sl Ann. Phys.} {\bf 253} (1997) 1.

\item{[2]} N. C. Tsamis and R. P. Woodard, ``The Quantum Gravitationally
Induced Stress Tensor During Inflation,'' {\sl hep-ph/9710444}.

\item{[3]} N. C. Tsamis and R. P. Woodard, {\sl Class. Quantum Grav.} 
{\bf 11} (1994) 2969.
\hfill\break
N. C. Tsamis and R. P. Woodard, {\sl Ann. Phys.} {\bf 238} (1995) 1.

\item{[4]} A. D. Dolgov, M. B. Einhorn and V. I. Zakharov, 
{\sl Phys. Rev.} {\bf D52} (1995) 717.  
 
\item{[5]} S. Wolfram, {\sl Mathematica} (Addison-Wesley, Redwood City,
Ca.).

\bye